%% file: main.tex
\documentclass[ALICE,manyauthors]{cernphprep}

\usepackage {amsmath}
\usepackage {amssymb}
\usepackage {graphicx}
\usepackage {subfigure}
\usepackage {bm}
\usepackage {indentfirst} %indent the first par after section
\usepackage{setspace}
\usepackage{color}
\usepackage{hyperref}

\usepackage{cite}
\usepackage{multirow}
\usepackage{datetime}
\usdate
\usepackage[toc,page]{appendix}
\usepackage{placeins}

\RequirePackage{lineno}
\usepackage[T1]{fontenc} 
\usepackage{orcidlink}

\input{./Content/Common/Include}

\begin{document}

\begin{titlepage}
\PHyear{2023}       % required, will be obtained from CERN
\PHnumber{180}      % required, will be obtained from CERN
\PHdate{24 August}  % required, will be obtained from CERN 

%\title{Search for jet quenching effects in high-multiplicity \pp\ collisions at $\sqrts=13$\,TeV via di-jet acoplanarity }

\title{Search for jet quenching effects in high-multiplicity \pp\ collisions at $\mathbf{\sqrt{\textit{s}} =13}$ TeV via di-jet acoplanarity }
\ShortTitle{Search for jet quenching effects in HM \pp\ collisions at $\sqrt{\textit{s}} =13$ TeV} 

\Collaboration{ALICE Collaboration\thanks{See Appendix~\ref{app:collab} for the list of collaboration members}}
\ShortAuthor{ALICE Collaboration} % appears on right page headers, do not change

%=================================================================
\begin{abstract}

The ALICE Collaboration reports a search for jet quenching effects in high-multiplicity (HM) proton--proton collisions at \sqrts\ = 13 TeV, using the semi-inclusive azimuthal-difference distribution \dphi\ of charged-particle jets recoiling from a high transverse momentum (high-\pTtrig) trigger hadron. Jet quenching may broaden the \dphi\ distribution measured in HM events compared to that in minimum bias (MB) events. The measurement employs a \pTtrig-differential observable for data-driven suppression of the contribution of multiple partonic interactions, which is the dominant background. While azimuthal broadening is indeed observed in HM compared to MB events, similar broadening for HM events is observed for simulations based on the PYTHIA 8 Monte Carlo generator, which does not incorporate jet quenching. Detailed analysis of these data and simulations show that the azimuthal broadening is due to bias of the HM selection towards events with multiple jets in the final state. The identification of this bias has implications for all jet quenching searches where selection is made on the event activity.

\end{abstract}
%=================================================================

\end{titlepage}

\setcounter{page}{2} %please do not remove this line

%%%%%%%%%%%%%%%%%%%%%%%%%%%%%%%%
% begin main text
%%%%%%%%%%%%%%%%%%%%%%%%%%%%%%%%

\input{Content/introduction}

\input{Content/dataset}

\input{Content/VzeroDistr}

\input{Content/jet_reconstruction}

\input{Content/raw_data_and_observables}

\input{Content/data_correction}

\input{Content/systematics}

\input{Content/Results}

%===================================================================
%%%%%%%%%%%%%%%%%%%%%%%%%%%%%%%%
% end main text 
%%%%%%%%%%%%%%%%%%%%%%%%%%%%%%%%

%%%%% acknowledgements - handled by EB chairs 
\newenvironment{acknowledgement}{\relax}{\relax}
\begin{acknowledgement}
\section*{Acknowledgements}
% add specific acknowledgements here 
% ...but please don't remove the line below: funding agencies
% will be acknowledged with a custom tex file handled by EB chairs after Collab Round 2
\input{fa_2023-08-02_Opt_C.tex}
\end{acknowledgement}

%%%%%%%% Bibliography 
\FloatBarrier
\bibliographystyle{utphys}
%\bibliography{../../Common/references}
\bibliography{./Content/Common/references}

%======================================================================

%%%%%%%%%%%%%%%%%%%%%%%%%%%%%%%%
% Appendices: yours (if any) + authorlist
%%%%%%%%%%%%%%%%%%%%%%%%%%%%%%%%
\newpage
\appendix

%%%%% Authorlist - please do not touch: handled by EB chairs 
\section{The ALICE Collaboration}
\label{app:collab}
\input{2023-08-02-Alice_Authorlist_2023-08-02_Opt_C.tex}
\end{document}

%% file: Content/Common/Include.tex
\newcommand{\CommentBlock}[1]{}

\newcommand{\VOMnorm}{\ensuremath{\text{V0M}/\langle\text{V0M}\rangle}}

\newcommand{\dAu}{d--Au}

\newcommand{\PbPb}{Pb--Pb}
\newcommand{\pPb}{p--Pb}

\newcommand{\pp}{pp}

\newcommand{\sqrtsNN}{\ensuremath{\sqrt{s_{\rm NN}}}}
\newcommand{\sqrts}{\ensuremath{\sqrt{s}}}
\newcommand{\rr}{\ensuremath{R}}

\newcommand{\gev}{\ensuremath{\mathrm{GeV/}c}}

\newcommand{\kT}{\ensuremath{k_\mathrm{T}}}
\newcommand{\antikT}{anti-\ensuremath{k_\mathrm{T}}}

\newcommand{\pT}{\ensuremath{p_\mathrm{T}}}

\newcommand{\pTjet}{\ensuremath{p_{\mathrm{T,jet}}}}
\newcommand{\pTjetch}{\ensuremath{p_\mathrm{T,jet}^\mathrm{ch}}}

\newcommand{\pTtrig}{\ensuremath{p_{\mathrm{T,trig}}}}

\newcommand{\ET}{\ensuremath{E_{\mathrm T}}}

\newcommand{\gammadir}{\ensuremath{\gamma_\mathrm{dir}}}

\newcommand{\qsqr}{\ensuremath{Q^2}}

\newcommand{\pizero}{\ensuremath{\pi^0}}

\newcommand{\dNjetdpTdphi}{\ensuremath{\frac{{\rm d}^{2}N_{\rm jet}}{\mathrm{d}\pTjetch\mathrm{d}\dphi}}}

\newcommand{\etajet}{\ensuremath{\eta_\mathrm{jet}}}

\newcommand{\qhat}{\ensuremath{\hat{q}}}
\newcommand{\vtwo}{\ensuremath{v_2}}

\newcommand{\Ntrig}{\ensuremath{N_{\rm{trig}}}}

\newcommand{\Drecoil}{\ensuremath{\Delta_\mathrm{recoil}}}
\newcommand{\DrecoilpTphi}{\ensuremath{\Delta_\mathrm{recoil}(\pT,\dphi)}}
\newcommand{\cRef}{\ensuremath{c_\mathrm{Ref}}}

\newcommand{\pTraw}{\ensuremath{p_\mathrm{T,jet}^\mathrm{raw,ch}}}
\newcommand{\pTrawi}{\ensuremath{p_\mathrm{T,jet}^\mathrm{raw,ch, i}}}

\newcommand{\pTreco}{\ensuremath{p_\mathrm{T,jet}^\mathrm{reco,ch}}}

\newcommand{\Ajet}{\ensuremath{A_\mathrm{jet}}}
\newcommand{\rhoA}{\ensuremath{\rho\times\Ajet}}

\newcommand{\Ajeti}{\ensuremath{A_\mathrm{jet}^\mathrm{i}}}

\newcommand{\dphi}{\ensuremath{\Delta\varphi}}

\newcommand{\TTSig}{\ensuremath{\mathrm{TT}_{\mathrm{Sig}}}}
\newcommand{\TTRef}{\ensuremath{\mathrm{TT}_{\mathrm{Ref}}}}

\newcommand{\zvtx}{\ensuremath{z_\mathrm{vtx}}}

\newcommand{\vzeroM}{\ensuremath{\mathrm{V0M}}}

\newcommand{\vzeroMscaled}{\ensuremath{\mathrm{V0M}/\left<\mathrm{V0M}\right>}}

\newcommand{\TT}[2]{\ensuremath{\mathrm{TT}\{#1,#2\}}}

\newcommand{\Rbkgd}{\ensuremath{R_{\mathrm{bkgd}}}}
\newcommand{\Rinstr}{\ensuremath{R_{\mathrm{instr}}}}

\newcommand{\Drecoilmeas}{\ensuremath{\Delta_\mathrm{recoil}^\mathrm{Meas}}}
\newcommand{\Drecoiltrue}{\ensuremath{\Delta_\mathrm{recoil}^\mathrm{True}}}

\newcommand{\dphidet}{\ensuremath{{\dphi}^\mathrm{det}}}
\newcommand{\dphipart}{\ensuremath{{\dphi}^\mathrm{part}}}
\newcommand{\pTjetdet}{\ensuremath{p_\mathrm{T,jet}^\mathrm{det}}}
\newcommand{\pTjetpart}{\ensuremath{p_\mathrm{T,jet}^\mathrm{part}}}

\newcommand{\argsdet}{\ensuremath{\left(\dphidet,\pTjetdet\right)}}
\newcommand{\argspart}{\ensuremath{\left(\dphipart,\pTjetpart\right)}}
\newcommand{\argsdetpart}{\ensuremath{\left(\dphidet,\pTjetdet;\dphipart,\pTjetpart\right)}}

%% file: Content/introduction.tex
%\newpage
\section{Introduction}
\label{Sect:Intro}

Strongly-interacting matter at extreme temperature and density forms quark--gluon plasma (QGP), a state in which the dominant degrees of freedom are sub-hadronic~\cite{Busza:2018rrf,ALICE:2022wpn}. The QGP filled the early universe, and it is generated today in the collision of heavy atomic nuclei at the Relativistic Heavy Ion Collider (RHIC) at Brookhaven National Laboratory and at the Large Hadron Collider (LHC) at CERN. Measurements at these facilities, and their comparison with theoretical calculations based on viscous relativistic hydrodynamics,
show that the QGP exhibits complex collective behavior, flowing with very low specific shear viscosity~\cite{Heinz:2013th}.

A key question in the experimental study of the QGP is the limit of its formation in terms of the size of the colliding nuclei. To explore this question, measurements have been carried out for ``small collision systems'', in which one of the projectiles is a proton or light nucleus and the other a heavy nucleus, additionally with selection of high event activity (EA) such as produced particle multiplicity or forward neutron production (see e.g. Ref.~\cite{Abelev:2015cnt}). Final-state hadronic distributions in small systems exhibit experimental signatures which are associated with production of the QGP in heavy-ion collisions~\cite{Nagle:2018nvi}, including collective flow~\cite{Khachatryan:2013ppb,Abelev:2013ppb,PHENIX:2018lia,STAR:2022pfn} and enhancement in the production of strange hadrons~\cite{Adam:2015strang}.

Jets are the hadronic remnants of hard (high momentum transfer \qsqr) interactions of quarks and gluons from the hadronic projectiles. Measurements of jet production and jet structure in elementary collisions are well described by theoretical calculations based on perturbative QCD (pQCD)~\cite{CMS:2016jip,ATLAS:2017ble,Acharya:2019jyg}. In heavy-ion collisions, jets interact with the QGP, resulting in modifications of jet production rates, structure, and angular distributions, which provide incisive probes of the QGP (``jet quenching'')~\cite{Majumder:2010qh,Cunqueiro:2021wls}. 

Jet quenching is a necessary consequence of QGP formation in small systems, though its effects are expected to be small~\cite{Zhang:2013oca,Tywoniuk:2014hta,Park:2016jap,Huss:2020whe, Huss:2020dwe,Zakharov:2021uza,Ke:2022gkq}. A common signature of jet quenching is the suppression of inclusive hadron or jet yield measured in heavy-ion collisions compared to that expected by scaling the corresponding yield measured in minimum-bias (MB) \pp\ collisions, using a Glauber modeling of the collision geometry~\cite{Miller:2007ri,Majumder:2010qh,Cunqueiro:2021wls}. Inclusive jet yield suppression has been reported using this approach in EA-selected \dAu\ collisions at RHIC (EA based on forward multiplicity)~\cite{Adare:2016dau} and in \pPb\ collisions at the LHC (EA based on forward transverse energy \ET)~\cite{ATLAS:2014cpa}. However, Glauber modeling using these EA metrics in small systems is subject to significant non-geometric 
bias~\cite{Adare:2013nff,Abelev:2015cnt,
Alvioli:2014eda, Bzdak:2014rca, Kordell:2016njg, Loizides:2017sqq, Alvioli:2017wou,Ke:2022gkq,PHENIX:2023dxl}. An alternative choice for the EA metric, based on zero-degree neutron measurements (nZDC), is found to be less biased, though the scaling still has model-dependent assumptions and uncertainties~\cite{Abelev:2015cnt}. Such biases and uncertainties limit the sensitivity of the measurement of jet quenching effects in small systems using Glauber-scaled inclusive yield observables. At present there is no significant evidence, beyond experimental uncertainties, of jet quenching in small systems using this approach~\cite{ALICE:2019fhe,ALICE:2021est,ALICE:2021wct}.

The PHENIX Collaboration has recently searched for jet quenching in \dAu\ collisions at a center-of-mass energy per nucleon--nucleon collision $\sqrtsNN=200$~GeV by the measurement of both \pizero\ and direct photon (\gammadir) inclusive yields in collisions selected by EA~\cite{PHENIX:2023dxl}. Jet quenching is studied using the \gammadir\ inclusive yield to estimate empirically the rate of hard processes, which does not depend upon Glauber scaling. Strong suppression in the ratio of \pizero\ and \gammadir\ inclusive yields is observed for high-EA relative to MB collisions, though with absolute value that is consistent with unity within the normalization uncertainty. 

An alternative approach has been proposed to search for medium-induced inclusive yield suppression in small systems utilizing MB collisions of light nuclei~\cite{Huss:2020dwe,Huss:2020whe}, which likewise does not require Glauber modeling for the yield scaling. However, this approach cannot be applied to EA-selected event populations in small collision systems, where the strongest signals characteristic of collective flow have been observed~\cite{Khachatryan:2013ppb,Abelev:2013ppb,PHENIX:2018lia,STAR:2022pfn,Adam:2015strang}. 

The comprehensive search for jet quenching effects in small collision systems therefore also requires approaches based on coincidence observables, which are self-normalized and likewise do not require a Glauber model calculation of nuclear geometry for an EA-selected population. The measurement of jets recoiling from a high-\pT\ hadron trigger in \pPb\ collisions at $\sqrtsNN=5.02$ TeV has set a limit of 0.4 \gev\ (90\% confidence) for medium-induced energy transport to angles greater than 0.4 radians relative to the jet axis, for high-EA collisions selected with criteria based both on forward multiplicity and on nZDC~\cite{Acharya:2017okq}. Measurement of the distribution of hadrons recoiling from a high-\pT\ jet trigger in EA-selected (nZDC) \pPb\ collisions at $\sqrtsNN=5.02$ TeV likewise finds no significant jet quenching signal within uncertainties~\cite{ATLAS:2022iyq}. The measurement of azimuthal anisotropy of high-\pT\ hadrons finds small but non-zero second Fourier coefficient \vtwo\ for events selected by EA based on forward \ET~\cite{ATLAS:2019vcm}, though such effects cannot be attributed solely to jet quenching. 

It therefore remains an open question whether the collective effects observed in small systems are indeed due to QGP formation, or whether they arise from other phenomena~\cite{Yan:2013laa,Bierlich:2017vhg, Blok:2017pui,Ke:2022gkq}. New searches for jet quenching effects in small systems are required to resolve this issue.

In this article we present a novel search for jet quenching effects in EA-selected high-multiplicity (HM) \pp\ collisions at $\sqrts=13$ TeV. Since ``collision geometry'' is ill-defined for EA-selected \pp\ collisions, inclusive observables are not appropriate for such a search. Rather, we utilize the semi-inclusive hadron+jet acoplanarity observable~\cite{Adam:2015pbpb,STAR:2017hhs,Acharya:2017okq,ALICE:2023qve,ALICE:2023jye}, i.e.~the distribution of the azimuthal angle \dphi\ between a high-\pT\ hadron trigger and correlated recoil jets, comparing \dphi\ measurements in HM-selected and MB populations. Jets are reconstructed from charged particles using the \antikT\ algorithm~\cite{FastJetAntikt} with resolution parameter $\rr=0.4$. Jet quenching in the QGP is expected to broaden the \dphi\ distribution relative to that in vacuum, due to in-medium multiple scattering~\cite{Appel:1985dq,Blaizot:1986ma,DEramo:2010wup,DEramo:2012uzl,Chen:2016vem,DEramo:2018eoy}. 
Indeed, significant in-medium acoplanarity broadening has been observed in central \PbPb\ collisions for large-\rr\ recoil jets at low \pT, though the physical mechanisms underlying this broaden remain an open question~\cite{ALICE:2023qve,ALICE:2023jye}. However, at present there is no theoretical guidance for the magnitude of the jet transport parameter \qhat~\cite{Majumder:2010qh} or alternative characterizations of jet quenching in EA-selected \pp\ collisions, and this is therefore entirely an experiment-driven search.

The analysis is based on the \Drecoil\ observable developed for semi-inclusive coincidence measurements of jets recoiling from a high-\pT\ hadron trigger~\cite{Adam:2015pbpb}. Precise suppression of 
jet yield uncorrelated with the trigger particle (uncorrelated background yield) is crucial in this analysis, since the uncorrelated yield of jets generated by multiple partonic interactions (MPI) from independent high-\qsqr\ processes can mimic azimuthal broadening arising from jet quenching. The \Drecoil\ observable provides data-driven suppression of uncorrelated background yield through the difference of trigger hadron-normalised recoil jet distributions in two exclusive trigger \pT\ intervals (Sec.~\ref{Sect:RawData_Observable}). Selection of the HM population utilizes a large data sample recorded by ALICE with an online HM trigger during the 2016--2018 LHC \pp\ runs at $\sqrts=13$ TeV. The \dphi\ distributions from the HM-selected and MB event populations are compared, revealing a striking acoplanarity broadening in HM-selected events. However, similar broadening is also observed in PYTHIA-based simulations. The physical origin of the broadening is elucidated through a detailed study of the rapidity dependence of jet production and the number of recoil jets in HM-selected and MB-selected events.

The paper is organized as follows: 
Sec.~\ref{Sect:Dataset} presents the data set and offline analysis; 
Sec.~\ref{Sect:VOMnormDistr} presents characterization of event activity using forward multiplicity; 
Sec.~\ref{Sect:Jet_reconstruction} presents jet reconstruction;
Sec.~\ref{Sect:RawData_Observable} presents the coincidence observable \Drecoil;
Sec.~\ref{Sect:pTcorr} presents data corrections;
Sec.~\ref{Sect:Systematics} presents systematic uncertainties;
and 
Sec.~\ref{Sect:Results} presents the physics results and their interpretation.

%% file: Content/dataset.tex
\section{Data set and offline analysis}
\label{Sect:Dataset}

The ALICE detector and its performance are described in Refs.~\cite{Aamodt:2008zz, Abelev:2014ffa}. Data for this analysis were recorded during the 2016, 2017, and 2018 LHC runs with pp collisions at $\sqrts=13$\,TeV. Events were selected online using signals in the V0 detectors~\cite{Abbas:2013vzr}, which are plastic scintillator arrays covering the pseudorapidity ranges $2.8 < \eta <5.1$ (V0A) and $-3.7<\eta<-1.7$ (V0C). The V0 signal is proportional to the total number of charged particles (multiplicity) in the detector acceptance. Two different V0 trigger configurations were employed, called minimum bias (labelled ``MB'') and high multiplicity (``HM''). The MB trigger required the in-time coincidence of V0A and V0C signals, while the HM trigger required the sum of V0A and V0C signal amplitudes (denoted as V0M) to be at least five times larger than the mean signal amplitude in MB events (denoted as $\left<{\rm V0M}\right>$). The HM trigger selected 0.1\% of MB events with the largest value of V0M. 

The EA is characterized offline by the scaled V0 signal, $\VOMnorm=(\text{V0A~+~V0C})/\left<\text{V0A~+~V0C}\right>$, which is insensitive to changes in V0 gain in the different data-taking periods due to scintillator aging. It also provides ordering of events in terms of EA without the need for precise calculation of the absolute V0 signal in model calculations, for a well-defined comparison of such models with data. The value of  $\left<{\rm V0M}\right>$ is calculated separately for each data-taking run lasting a few hours, as a function of the collision vertex position along the beam axis. The HM selection is further constrained in the offline analysis to the range $5<\VOMnorm<9$. The lower bound of 5 is determined by the online HM trigger threshold, while the upper bound of 9 is determined by the range over which the  \VOMnorm\ distributions for the three different measurement periods are consistent; higher values may be affected by residual, uncorrected pileup effects. 

In the offline analysis, jets are measured at midrapidity using charged particles reconstructed with the ALICE central barrel detectors, covering the range $|\eta|<0.9$. Track reconstruction is based on space points measured by the Inner Tracking System (ITS) and Time Projection Chamber (TPC)~\cite{Abelev:2014ffa}. Primary event vertices are reconstructed offline based on global tracks, which are required to have space points in the Silicon Pixel Detector (SPD) forming the two innermost layers of the ITS. Accepted events are required to have the primary vertex  within $\left|\zvtx \right| < 10$\,cm, where \zvtx\ is the location of the vertex along the beam axis relative to the nominal center of the ALICE detector.

For MB-triggered events, the pileup rate due to multiple hadronic \pp\ collisions in the same LHC bunch crossing is less than 3.5\%. The pile-up contribution is suppressed offline by rejecting events with multiple reconstructed event vertices. Monte Carlo studies estimate that the residual pileup contribution following this cut is negligible for the MB populations and about 1\%  for the HM population. As discussed in Sec.~\ref{Sect:RawData_Observable}, the observable \Drecoil\ used in the analysis provides data-driven suppression of uncorrelated background yield, which also includes residual pileup events that are not rejected by the multiple-vertex algorithm. After event selection, the data sets have an integrated luminosity of 32\,nb$^{-1}$ for the MB trigger and 10\,pb$^{-1}$ for the HM trigger.

During the data taking, the ITS had non-uniform efficiency, and the analysis therefore utilizes hybrid tracks~\cite{Acharya:2019pp5,Acharya:2019jyg} to achieve azimuthally uniform tracking response. Hybrid tracks consist of good quality global tracks with at least one hit in the SPD, and complementary tracks without SPD signals. To ensure good momentum resolution, the momentum of these complementary tracks is determined using the primary vertex as a constraint. Reconstructed tracks with $|\eta|<0.9$ and $\pT>0.15$ \gev\ are accepted for the analysis.  Hybrid track reconstruction efficiency is 0.85 at $\pT = 1$\,\gev, 0.82 at $\pT = 10$\,\gev, and 0.74 at $\pT = 50$\,\gev. Tracking efficiencies for MB and HM events are similar.
Primary-track momentum resolution is
0.7\% at  $\pT = 1$\,\gev, 1.3\% at $\pT = 10$\,\gev, and 3.7\% at $\pT = 50$\,\gev.

Simulations are utilized for data corrections and for comparison to theoretical calculations. The simulations are based on the PYTHIA 8 event generator~\cite{Pythia8:2007} with Monash tune~\cite{Skands:2014pea}, and a detailed GEANT3 model~\cite{Brun:1119728} of the ALICE detector response, which includes production of secondary particles and realistic hit digitization. Events generated by PYTHIA 8 without detector effects are denoted ``particle-level,'' and such events passed through GEANT3 are denoted ``detector-level.''

For particle-level events, the V0A and V0C responses are determined by counting the number of charged particles in their acceptance.  The coincidence requirement of the online trigger is modeled by requiring particle-level events to have particles in both V0A and V0C, while detector-level events are required to have GEANT-generated hits in both V0A and V0C. The MB events are used to calculate the \vzeroM\ distributions at both detector and particle level.

%% file: Content/VzeroDistr.tex
\section{\VOMnorm\ distributions}
\label{Sect:VOMnormDistr}

%----
\begin{figure}[htb]
\centerline{%
\includegraphics[width=0.8\textwidth]{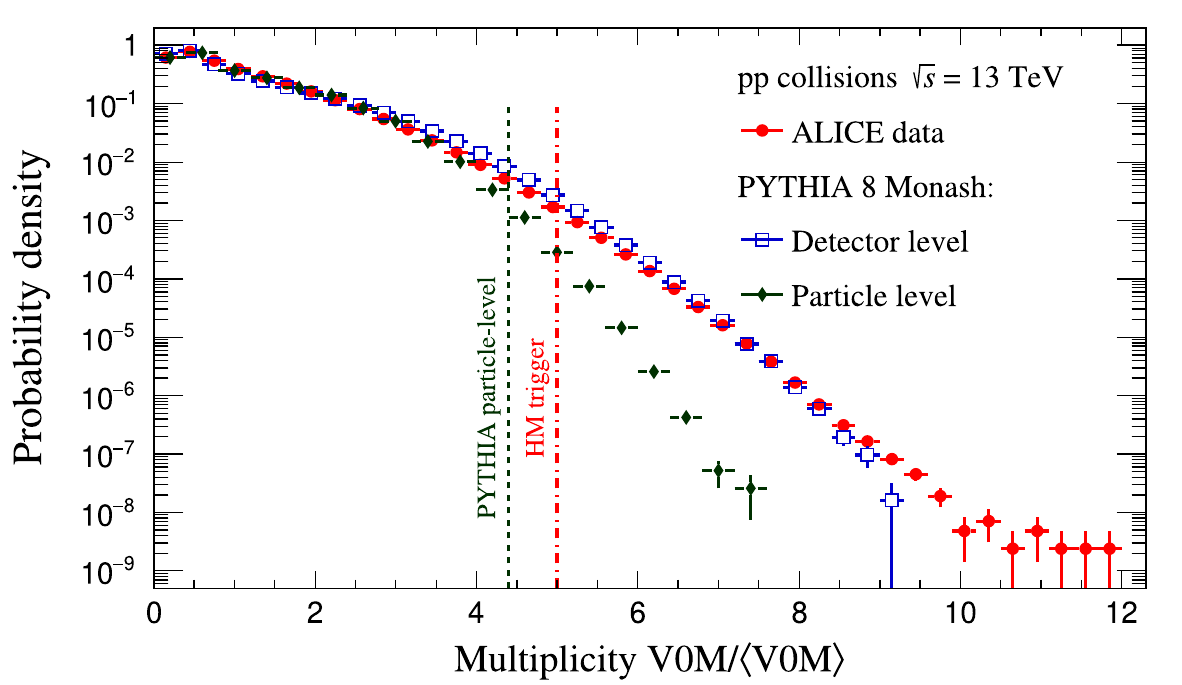}}
\caption{Probability distribution of \VOMnorm\ in MB pp collisions measured at $\sqrt{s}=13$\,TeV, and in simulated MB pp events generated by PYTHIA 8 at the particle and detector level. The vertical dashed lines indicate the lower bound for HM selection for data and particle-level simulations.}
\label{fig:V0Mnorm}
\end{figure}
%----

Figure~\ref{fig:V0Mnorm} shows the \VOMnorm\ probability distribution for MB \pp\ collisions at $\sqrts=13$\,TeV. The lower limit for HM event selection, $\VOMnorm=5$, is indicated by the red dashed-dotted line. The figure also shows the \VOMnorm\ probability distribution at the particle and detector level for PYTHIA~8-generated events for MB \pp\ collisions at $\sqrts=13$\,TeV. These distributions differ because of a large contribution at forward angles of secondary particles generated in detector material~\cite{ALICE:2004ftm}.  The particle-level distribution falls more rapidly than that observed in data in the range $\VOMnorm>4$. However, the detector-level distribution qualitatively reproduces that observed in data, with probability densities which lie within a factor $\sim2$ of each other over a range of nine orders of magnitude in \VOMnorm.

In order to compare PYTHIA~8 particle-level HM-selected distributions to data, we assume that the secondary particle yield due to interactions in detector material is on average proportional to the primary multiplicity in the V0 acceptance. The HM selection for the particle-level distribution is therefore chosen to select the same fraction of the MB cross section (0.1\%) as the HM selection $\VOMnorm>5.0$ used for data. This selection corresponds to particle-level $\VOMnorm>4.4$,  as indicated by the dark dashed line in Fig.~\ref{fig:V0Mnorm}.

We note, however, that such selections of the same cross section fraction at the particle and detector level only select the same fraction of the distributions. They cannot select the same population event-by-event, due to significant fluctuations in the correlation between particle- and detector-level events. For the detector model used in this study, about 35\% of events passing the HM selection at the  detector level  also do so at the particle level. This is a generic issue, with similar features expected for any event selection in small systems based on forward EA.

%% file: Content/jet_reconstruction.tex
\section{Jet reconstruction}
\label{Sect:Jet_reconstruction}

Jet measurements in this coincidence analysis are corrected for two distinct background effects: uncorrelated jet yield, which is corrected using the \Drecoil\ observable discussed in Sect.~\ref{Sect:RawData_Observable}; and \pT-smearing of correlated jets due to overlap with uncorrelated background components, as detailed in this section. These corrections are carried out in distinct analysis steps.

Several types of reconstructed jets are consequently used in the analysis, which are distinguished using the notation defined in Refs.~\cite{Adam:2015pbpb,Acharya:2017okq}.
For data, \pTraw\ refers to the raw output of the jet reconstruction algorithm; \pTreco\ refers to \pTraw\ after subtraction of the event-wise estimate of the background contribution to \pTjet, \rhoA, which is the product of median jet \pT\ density in the event and the jet area (Eq.~\ref{eq:rho}); and \pTjetch\ refers to the fully corrected jet transverse momentum. For simulations, \pTjetpart\ refers to charged-particle jets at the particle level, and \pTjetdet\ charged-particle jets at the detector level; both quantities are corrected by \rhoA\ using the following procedure.

Jets are reconstructed from accepted charged-particle tracks. Particles are assumed to be massless and their four-momenta are combined with the boost-invariant \pT\ recombination scheme~\cite{Cacciari:2011ma}.  Jet reconstruction is carried out twice for each event. The first reconstruction pass uses the \kT\ algorithm~\cite{Cacciari:2011ma} with $\rr=0.4$ and accepts jets with $|\etajet|<0.9-\rr$. The first-pass jet population is used to determine $\rho$, the event-wise estimate of the background energy density~\cite{Cacciari:2007fd},

%----
\begin{equation}
\rho = \mathrm{median} \left\lbrace \frac{\pTrawi}{\Ajeti} \right\rbrace,
\label{eq:rho}
\end{equation}
%----

\noindent
where \pTrawi\ and \Ajeti\ are the raw jet \pT\ and the area~\cite{Cacciari:2008gn} of the $i^\mathrm{th}$ jet in the event. Jet area is calculated using the ghost area method of FastJet, with a ghost area of 0.005~\cite{Cacciari:2008gn}. The two hardest jets in the event are excluded from the median calculation. The most probable value of $\rho$ is zero in both the MB and HM populations, while the mean value of $\rho$ is 0.09~\gev\ for MB and 1.24~\gev\ for HM-selected events. For events containing a charged track in $|\eta|<0.9$ with $20<\pT<30$~\gev, the mean value of $\rho$ is 0.39~\gev\ for MB and 1.62~\gev\ for HM events.

The second reconstruction pass uses the \antikT\ algorithm~\cite{Cacciari:2011ma} with $\rr=0.4$. The acceptance for the second pass is likewise $|\etajet|<0.9-\rr$ over the full azimuth. 

The jet \pT\ obtained from the second pass is then adjusted for the  median background \pT\ density $\rho$ according to~\cite{Cacciari:2007fd}

\begin{equation}
\pTreco = \pTraw - \rhoA.
\label{Eq:pT_Correction}
\end{equation}

\noindent
The jet \pT\ scale and jet \pT\ resolution are the same as in Ref.~\cite{ALICE:2022jbp}.

%% file: Content/raw_data_and_observables.tex
%\FloatBarrier
\section{Observables and raw data}
\label{Sect:RawData_Observable}

The analysis utilizes a differential observable based on the semi-inclusive distribution of charged-particle jets recoiling from a high-\pT\ trigger (``h+jet'')~\cite{Adam:2015pbpb} (see also~\cite{STAR:2017hhs,Acharya:2017okq}). The key components of this approach are summarized in this section.

The goal of the analysis is the search for broadening of the \dphi\ distribution in HM-selected events due to medium-induced jet scattering, by comparison to the MB population. A significant source of uncorrelated background yield to this process arises from MPIs, in which multiple uncorrelated high-\qsqr\ partonic interactions occur in the same \pp\ collision, with one such interaction generating a trigger hadron and another generating a recoil jet in the acceptance. The \dphi\ distribution of such MPI pairs is by definition uniform on average, thereby limiting the measurement sensitivity to broadening of the \dphi\ distribution from medium-induced scattering of correlated recoil jets.

Precise background yield correction must be carried out in a fully data-driven way, without model dependence. We therefore employ the \Drecoil\ observable~\cite{Adam:2015pbpb}, which is the difference between semi-inclusive recoil jet distributions for two ranges of \pTtrig, both normalized to the number of trigger hadrons,

%----
\begin{equation}
\DrecoilpTphi =
\frac{1}{\Ntrig}\dNjetdpTdphi\Bigg\vert_{\pTtrig\in{\TTSig}}  -
\cRef\times \frac{1}{\Ntrig}\dNjetdpTdphi\Bigg\vert_{\pTtrig\in{\TTRef}},
\label{eq:DRecoil}
\end{equation}
%----

\noindent 
where TT denotes ``trigger track.'' In this analysis, $\TTSig=\TT{20}{30}$ specifies the range $20<\pTtrig<30$~\gev\ for the Signal trigger distribution, and $\TTRef=\TT{6}{7}$ specifies the range $6<\pTtrig<7$~\gev\ for the Reference trigger distribution. These intervals were chosen to optimize the opposing requirements of obtaining high statistical precision and limiting the kinematic range for more precise comparison of the same observables with different EA selections. The number of trigger hadrons measured in each TT class is denoted \Ntrig. The azimuthal difference \dphi\ between TT and recoil jet is defined to have the range $[0,\pi]$~radians.

\begin{figure}[htb]
\centering
\includegraphics[width=1.0\textwidth]{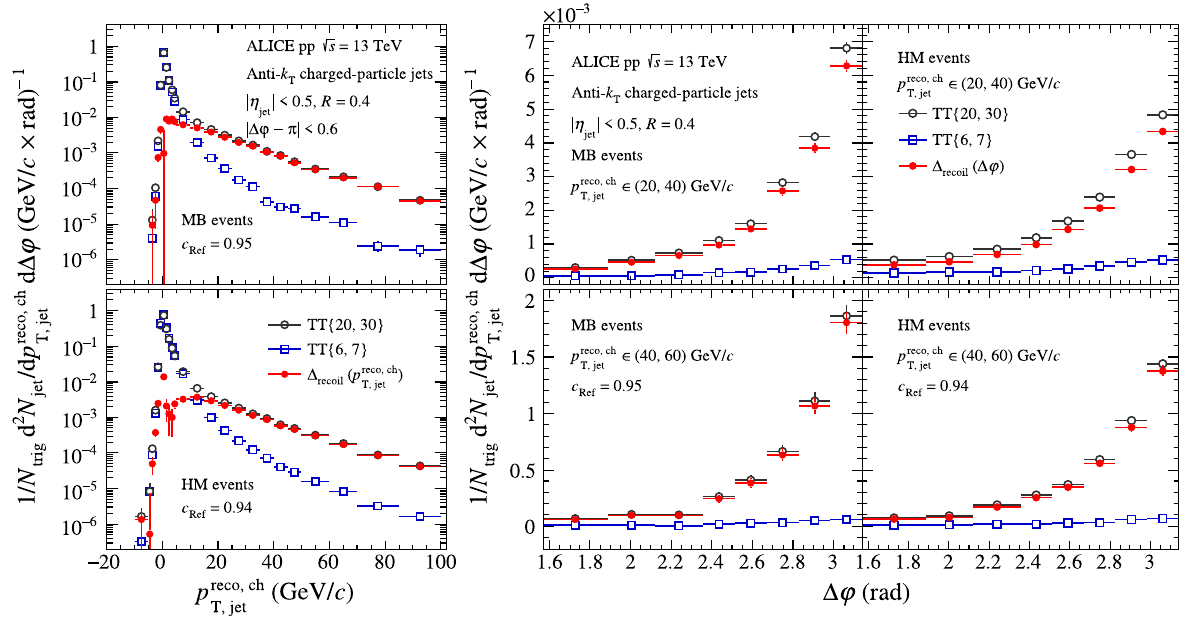}
\caption{The Signal (\TT{20}{30}) and Reference (\TT{6}{7}) trigger-normalized recoil jet distributions and the corresponding \Drecoil\ distribution for MB and HM \pp\ collisions at $\sqrts = 13$ TeV. Left panels: projection onto \pTreco\ for $|\dphi-\pi|<0.6$. Middle and right panels: projection onto \dphi\ in two \pTreco\ intervals.}
\label{fig:RawDrecoilSpectraAco}
\end{figure}

The \Drecoil\ distribution is a two-dimensional function of \pTjet\ and \dphi~\cite{Adam:2015pbpb,Acharya:2017okq}. We define its one-dimensional projections, \Drecoil(\pTjet) and \Drecoil(\dphi), onto the \pTreco\ and \dphi\ axes respectively, for restricted ranges in the other kinematic variable. These projections are shown in Figs.~\ref{fig:RawDrecoilSpectraAco} and~\ref{fig:DrecoilSpectraAco} both prior to and after corrections, as indicated by the functional argument (e.g. \Drecoil(\pTreco) or \Drecoil(\pTjetch)).

The scaling factor \cRef, which is applied to the Reference distribution (second term in Eq.~\ref{eq:DRecoil}), accounts for the different phase space available to observe uncorrelated yield in the Signal and Reference distributions~\cite{Adam:2015pbpb,STAR:2017hhs}. In this analysis, the value of \cRef\ is determined from the ratio of trigger-normalized Signal and Reference recoil jet yields in the bin $0<\pTreco<1$~\gev, which is expected to be dominated by uncorrelated background yield. This gives values $\cRef=0.95\pm{0.03}~\text{(syst.)}$ in MB events and $\cRef=0.94\pm{0.03}~\text{(syst.)}$ in HM events.

Since the uncorrelated yield is by definition independent of TT, it therefore contributes with equal magnitude to the two terms in Eq.~\ref{eq:DRecoil} and is therefore removed by the subtraction. While \Drecoil\ is a differential observable and not an absolutely normalized yield, its two terms are nevertheless calculable perturbatively~\cite{deFlorian:2009fw}. Measurements of \Drecoil\ in minimum-bias \pp\ collisions are well described by PYTHIA 8~\cite{Adam:2015pbpb,Acharya:2017okq}.

In the analysis of both MB and HM-selected events, the dataset is divided into two statistically independent subsets, with the Signal distribution determined using 95\% of all events, and the Reference distribution determined using the remaining 5\%. These fractions were chosen to provide an equal number of trigger hadrons in the two TT classes, in order to optimize the statistical precision of \Drecoil. The statistical error due to the \Ntrig\ normalization is negligible.

If two hadrons in an event satisfy the TT condition, one is chosen at random. This selection ensures that the \pT-differential TT distribution has the same shape as the inclusive charged-hadron yield, which is an essential requirement for a semi-inclusive measurement~\cite{Adam:2015pbpb}.

%--------

%-------

Figure~\ref{fig:RawDrecoilSpectraAco} shows selected \Drecoil(\pTreco) and \Drecoil(\dphi) distributions, together with their corresponding Signal and Reference  distributions. The Signal and Reference distributions have similar magnitude only in the region $\pTreco<20$~\gev, while at larger values of \pTreco\ the Reference distribution falls below the Signal distribution and \Drecoil\ is similar in magnitude to the Signal distribution.

%% file: Content/data_correction.tex
%\FloatBarrier
\section{Corrections}
\label{Sect:pTcorr}

Particle-level jets are clustered from all final-state charged particles generated with PYTHIA 8 as described in Sec.~\ref{Sect:Dataset}, except for weak decay daughters~\cite{Acharya:2017def,Abelev:2013kqa}. The measured (detector-level) \Drecoil\ distribution is corrected to the particle level. 

Subtraction of the scaled Reference distribution in Eq.~\ref{eq:DRecoil} accurately removes the uncorrelated jet yield~\cite{Adam:2015pbpb}. However, the resulting \Drecoil\ distribution is still subject to instrumental effects which smear \pTjet\ and \dphi, whose correction is discussed in this section. These smearing effects are encoded in the four-dimensional response matrix \Rinstr, which maps the true (particle-level) distribution of \Drecoil\ onto the measured distribution,

\begin{equation}
\Drecoilmeas\argsdet = \Rinstr\argsdetpart \otimes \Drecoiltrue\argspart.
\label{Eq:Convolution_TrueDistrib}
\end{equation}

The \pTjet\ and angular smearing due to instrumental effects is very similar in the inclusive and recoil jet populations in these simulations. In addition, the inclusive population has a substantially larger sample in the simulated dataset. The matrix \Rinstr\ is therefore constructed using the inclusive jet population, by matching particle-level and detector-level jets in phase space within $\Delta\rr = \sqrt{(\Delta\eta)^{2} + (\dphi)^{2}} < 0.3$. 
The response matrix has \pT\ bins of width 1~\gev, so that the difference in the spectrum shape of the inclusive and recoil jet populations has negligible effect on this procedure.

Track reconstruction efficiency, which is the dominant instrumental effect, is found to be the same for MB and HM events. The MB and HM analyses therefore use the same response matrix, obtained using MB events. Unmatched particle-level jets are tabulated and their rate is applied as an efficiency correction, following the unfolding correction discussed below. Jet matching efficiency for both MB and HM events  is 0.98 at $\pTjetch=10$~\gev\ and consistent with unity at higher \pTjetch. 

The corrected \Drecoil\ distribution is calculated by regularized inversion of Eq.~\ref{Eq:Convolution_TrueDistrib}, using two-dimensional iterative Bayesian unfolding~\cite{dAgostini2010} implemented in the RooUnfold package~\cite{RooUnfold}. The input distribution \Drecoilmeas\ is specified in the range $10<\pTjetdet<100$~\gev\ and $\pi/2<\dphidet<\pi$ rad. The prior distribution for initiating the unfolding is the \Drecoil\ particle-level spectrum calculated using PYTHIA 8 simulations. For unfolding the MB dataset, the prior is calculated using MB PYTHIA 8 events, whereas for unfolding the HM dataset the detector-level HM selection is applied to the simulated data, as shown in Fig.~\ref{fig:V0Mnorm}.

Regularization is optimized by requiring that the unfolded distributions from successive iterations exhibit a mean change of less than 2\%, averaged over all \pT\ bins. The optimum number of iterations using this criterion is found to be 5, for both the MB and HM analyses.

While correction of the \Drecoil\ distribution for instrumental effects is carried out by regularized unfolding using the full instrumental response matrix \Rinstr, insight into the unfolding procedure can also be gained by parametric characterization of the main detector-level effects. Detector-level effects in \pTjet\ are characterized by the jet energy scale shift (JES) between particle-level and detector-level, $\langle(\pTjetdet-\pTjetpart)/\pTjetpart \rangle$, while jet energy resolution (JER) is its width, $\sigma (\pTjetdet)/\pTjetpart$, where $\sigma (\pTjetdet)$ is RMS of the \pTjetdet\ distribution. For $\pTjetpart=10,~40,~\text{and}~70$ \gev, JES is $-10\%, -14\%,~\text{and}~-18\%$, and JER is $25\%,~22\%,~\text{and}~23\%$, respectively. Other systematic effects which are not encoded in the response matrix, such as the precision of the ALICE central barrel $B$-field scale ($\sim10^{-3}$), make negligible contribution to JES and JER.

Azimuthal angular smearing is likewise found to have the same magnitude in MB and HM events. Azimuthal smearing of tracks contributing to \TT{20}{30} has $\text{RMS}\approx 2 \times 10^{-3}$~rad, while azimuthal smearing of the jet axis has $\text{RMS}\approx 26 \times 10^{-3}$~rad for $10<\pTjetch<15$~\gev\ and $11 \times 10^{-3}$~rad for $60<\pTjetch<80$~\gev. The corresponding distributions of \dphi\ have RMS of $34 \times 10^{-3}$~rad for $10<\pTjetch<15$~\gev\ and $13 \times 10^{-3}$~rad for $60<\pTjetch<80$~\gev.

Subtraction of the reference distribution in the \Drecoil\ observable corrects the uncorrelated background yield. However, the resulting two-dimensional distribution as a function of \pTjetch\ and \dphi\ is still smeared by residual fluctuations of the uncorrelated background component, which causes the underlying event density to deviate locally from $\rho$. Correction for such fluctuations requires model assumptions (see e.g. Refs.~\cite{ALICE:2022jbp,ALICE:2021wct}) and is therefore not included in the unfolding; rather, its magnitude  is assessed for two common model choices and the variation in unfolded distributions is included in the systematic uncertainty, as discussed in Sec.~\ref{sect:sysuncertbkgd}.

The unfolding procedure is validated by a closure test in which the input is PYTHIA 8 detector-level events, the full analysis chain, including unfolding, is carried out, and the output is compared to the particle-level PYTHIA 8 spectrum. Good agreement between both distributions is found within statistical uncertainties, thereby confirming the robustness of the applied corrections.

%% file: Content/systematics.tex
%\FloatBarrier
\section{Systematic uncertainties}
\label{Sect:Systematics}

The main sources of systematic uncertainty in the measurement of \Drecoil\ are related to track reconstruction efficiency; track \pT\ resolution; the unfolding procedure; the determination of the scaling factor \cRef; and residual background \pT-density fluctuations relative to $\rho$. The systematic uncertainty due to each source is estimated by varying the appropriate parameters and rerunning the full analysis chain.  

However, the finite statistical precision of the data imposes a limit on the precision with which the spectrum can be unfolded. This limit is taken into account in the determination of the systematic uncertainties 
by using the following procedure, which is described in detail in Ref.~\cite{Acharya:2017okq}. For each parameter variation the spectrum is unfolded 10 times, with the central values of the spectrum varied randomly and independently using a Poisson distribution corresponding to the statistical error of each data point~\cite{Acharya:2017okq}. The ratio of the unfolded spectrum from each iteration to that of the primary analysis  is calculated, and at each point the median of this set of ratios is assigned as the systematic uncertainty from this source. Tables~\ref{tab:SystUncerMB} and~\ref{tab:SystUncerHM} show representative values of the systematic uncertainty in \Drecoil(\pTjetch) and \Drecoil(\dphi) measurements for MB and HM-selected event populations, respectively.

%----------------
\begin{table}[!hbt]
\centering
\small
\caption{Main sources of systematic uncertainty and total uncertainty in \Drecoil(\pTjetch) and \Drecoil(\dphi) in representative bins, for MB events.}
\begin{tabularx}{1.0\textwidth}{c|c|c|c|c|c|c}
\hline
  & \multicolumn{6}{c}{\bf Relative systematic uncertainty (\%)} \\ [0.7ex]
\hline
 \multirow{2}{*}{Projection} & \multicolumn{2}{c|}{\Drecoil(\pTjetch)} & \multicolumn{2}{c|}{\Drecoil(\dphi)} & \multicolumn{2}{c}{\Drecoil(\dphi)} \\ [0.7ex]
  & \multicolumn{2}{c|}{$|\dphi - \pi| < 0.6$} & \multicolumn{2}{c|}{
  $\pTjetch \in (20,40)$\,\gev} & \multicolumn{2}{c}{$\pTjetch \in (40,60)$\,\gev} \\ [0.7ex]
\hline 
 Bin & 10--20\,\gev & 60--80\,\gev & $2\pi/3$ & $\pi$ & \multicolumn{1}{c|}{$2\pi/3$} & \multicolumn{1}{c}{$\pi$} \\ [0.7ex]
\hline
\hline
Tracking efficiency             & 0.2 & 7.1 & 3.7 & 1.9 & 8.4 & 5.6 \\ [0.7ex]
Track $p_{\text{T}}$ resolution & 0.3 & 0.3 & 0.2 & 0.2 & $\sim 0$ & 0.3 \\ [0.7ex]
Unfolding procedure             & 0.6 & 0.7 & 3.9 & 0.3 & 3.4 & 0.9 \\ [0.7ex]
\cRef\ variation      & $(-1.7, 1.9)$ & $(-0.2, 0.2)$ & $(-0.6, 0.7)$ & $(-0.4, 0.4)$ & $(-0.5, 0.5)$ & $(-0.1, 0.2)$  \\ [0.7ex]
Bkgd fluctuations              & $(-1.6, 0)$ & $(-2.4, 0)$ & $(-4.7, 0)$ & $(-1.6, 0)$ & $(-5.0, 0)$ & $(-2.7, 0)$ \\ [0.7ex]
\hline
Total uncertainty & $(-2.5,~2.0)$ & $(-7.6, 7.2)$ & $(-7.2, 5.4)$ & $(-2.5, 2.0)$ & $(-10.4, 9.0)$ & $(-6.3, 5.7)$ \\ [0.1ex]
\hline
\end{tabularx}
\label{tab:SystUncerMB}
\end{table}
%--------------------------

%-------------------------
\begin{table}[!hbt]
\centering
\small
\caption{Same as Tab.~\ref{tab:SystUncerMB}, for HM-selected events.}
\begin{tabularx}{1.0\textwidth}{c|c|c|c|c|c|c}
\hline
  & \multicolumn{6}{c}{\bf Relative systematic uncertainty (\%)} \\ [0.7ex]
\hline
 \multirow{2}{*}{Projection} & \multicolumn{2}{c|}{$\Drecoil (\pTjetch)$} & \multicolumn{2}{c|}{$\Drecoil (\dphi)$} & \multicolumn{2}{c}{$\Drecoil (\dphi)$} \\ [0.7ex]
  & \multicolumn{2}{c|}{$|\dphi - \pi| < 0.6$} & \multicolumn{2}{c|}{$\pTjetch \in (20,40)$\,\gev} & \multicolumn{2}{c}{$\pTjetch \in (40,60)$\,\gev} \\ [0.7ex]
\hline 
 Bin & 10--20\,\gev & 60--80\,\gev & $2\pi/3$ & $\pi$ & \multicolumn{1}{c|}{$2\pi/3$} & \multicolumn{1}{c}{$\pi$} \\ [0.7ex]
\hline
\hline
Tracking efficiency             & 0.6 & 7.6 & 4.2 & 1.9 & 7.1 & 5.2 \\ [0.7ex]
Track $p_{\text{T}}$ resolution & 0.3 & 0.1 & 0.4 & 0.2 & 0.1 & 0.3  \\ [0.7ex]
Unfolding procedure             & 0.9 & 0.7 & 5.7 & 1.6 & 2.6 & 1.2 \\ [0.7ex]
\cRef\ variation      & $(-4.5, 3.2)$ & $(-0.3, 0.2)$ & $(-1.5, 1.1)$ & $(-0.6, 0.5)$ & $(-1.3, 0.9)$ & $(-0.3, 0.2)$  \\ [0.7ex]
Bkgd fluctuations              & $(-1.5, 0)$ & $(-3.0, 0)$ & $(-7.4, 0)$ & $(-2.1, 0)$ & $(-4.5, 0)$ & $(-3.2, 0)$ \\ [0.7ex]
\hline
Total uncertainty & $(-4.6, 3.7)$ & $(-8.2, 7.6)$ & $(-10.3, 7.2)$ & $(-3.3, 2.5)$ & $(-8.9, 7.6)$ & $(-6.3, 5.4)$ \\ [0.1ex]
\hline
\end{tabularx}
\label{tab:SystUncerHM}
\end{table}
%---------------------

%-------
\subsection{Tracking efficiency and track $\mathbf{\textit{p}_\mathrm{\bf T}}$ resolution}
\label{sect:sysuncerttrack}

The tracking efficiency uncertainty is 3\%~\cite{Acharya:2019pp5}. To assess the corresponding uncertainty in the \Drecoil\ distribution, a variation of \Rinstr\ is constructed in which 3\% of all tracks are randomly discarded. While it is in practice not possible to generate \Rinstr\ with 3\% higher tracking efficiency, this uncertainty is expected to be symmetric. The resulting uncertainty ranges from $<1$\% at low \pTjetch\ to 7\% at high \pTjetch, with only minor dependence on EA.

To assess the systematic uncertainty due to track \pT\ resolution, two different instances of \Rinstr\ are generated, with \pT-resolution corresponding to that of either real data or detector-level MC data hybrid tracks. This source makes negligible contribution to the total systematic uncertainty.

%-------
\subsection{Unfolding}
\label{sect:sysuncertunfold}

The unfolding procedure has several parameters whose values influence the corrected \Drecoil\ distribution: number of iterations; choice of prior spectrum; and range and binning of the raw input distribution. Each source was varied independently:

\begin{itemize}
\item The regularization condition was varied by $\pm 1$ iterations with respect to the optimized value of 5. The corresponding uncertainty is found to be small, since unfolding converges rapidly to a stable result. 

\item Variations in the prior spectrum were obtained using the particle-level \Drecoil\ spectra generated by the POWHEG MC event generator~\cite{Frixione:2007vw,Alioli:2011vw} matched to PYTHIA~8 for parton shower and hadronization, and with different choices of regularization and factorization scale.

\item The \pTjet\ binning was varied by shifting the bin boundaries by 1--2~\gev, and by changing the lower bound of the input spectrum from 10~\gev\ to 6~\gev. The binning in \dphi\ was not varied, since \dphi\ smearing effects are small.
\end{itemize}

The systematic uncertainty attributed to unfolding is the maximum deviation in the \Drecoil\ spectrum from varying these parameters, relative to the \Drecoil\ spectrum using the primary analysis parameters. For \Drecoil(\pTjetch), the resulting relative systematic uncertainty is about 0.6\% for MB and 0.9\% for HM events, with a weak dependence on \pTjetch. For \Drecoil(\dphi), the relative systematic uncertainty is smallest at $\dphi=\pi$ and increases monotonically towards $\dphi=\pi/2$, for both MB and HM events.

%-------
\subsection{Scaling factor $\mathbf{\textit{c}_\mathrm{\bf Ref}}$}
\label{sect:sysuncertcRef}

The value of the \cRef\ scaling factor in Eq.~\ref{eq:DRecoil} was varied in the range $[0.9, 1]$. This range brackets the \cRef\ values obtained by changing the \pTreco\ bin in which it is evaluated from $(0,1)$~\gev\ to $(-1,0)$~\gev, and by its variation with \dphi. Different choices of \cRef\ modify the \Drecoil\ spectrum relative to that of the primary analysis result, with uncertainty decreasing as a function of \pTjetch. Representative values are provided in Tables~\ref{tab:SystUncerMB} and~\ref{tab:SystUncerHM}.

%-------------------------------------------------
\subsection{Background fluctuations}
\label{sect:sysuncertbkgd}

As discussed in Sec.~\ref{Sect:pTcorr}, no correction is applied directly for the effect of residual background fluctuations; rather, a model-dependent estimate of its magnitude contributes to the systematic uncertainty. For this estimate, a response matrix which encodes the effect of residual background fluctuations, \Rbkgd, is convoluted with the instrumental response matrix \Rinstr\ in Eq.~\ref{Eq:Convolution_TrueDistrib}. The matrix \Rbkgd\ is determined for events selected with \TT{20}{30}, using two methods:

\begin{itemize}

\item Calculate the sum of track \pT\ in a cone \rr\    = 0.4 placed randomly in the acceptance,  excluding overlap with the leading and sub-leading jets, and the TT. This sum is corrected for the median background density, 
\begin{equation}
    \delta \pT^{\text{RC}} = \sum_{\text{tracks}~\in~\text{RC}} p_{\text{T, track}} - \rho \times \pi R^{2},
\end{equation}
where $\pi R^{2}$ is the cone area and $\rho$ is defined in Eq.~\ref{eq:rho}. The matrix \Rbkgd\ is the distribution of $\delta \pT^{\text{RC}}$.

\item Embed a high-\pT\ track perpendicular in azimuth to the TT and at the same value of $\eta$, with \pT\ distributed uniformly in the range 0--20~\gev. Jet reconstruction is then carried out, the jet candidate containing the embedded track is identified, and the quantity $\delta\pT^{\text{ET}}$  is calculated as
\begin{equation}
    \delta \pT^{\text{ET}} = p_{\text{T, jet}}^{\text{ch, emb}} - \rho \times A_{\text{jet}}^{\text{emb}} - \pT^{\text{emb}},
\end{equation}
where $\pT^{\text{emb}}$ is transverse momentum of the embedded track. The matrix \Rbkgd\ is the distribution of $\delta \pT^{\text{ET}}$.
\end{itemize}

\noindent
Unfolding is then carried out for both choices, and the assigned systematic uncertainty is the maximum difference of the two unfolded \Drecoil\ distributions from that of the primary analysis. 

%------------------------
\subsection{Total systematic uncertainty}
\label{sect:sysuncerttotal}

The total systematic uncertainty of the unfolded \Drecoil\ distribution is the quadrature sum of the contribution from each source. The systematic uncertainties from all sources except unfolding are correlated between the HM and MB analyses. This correlation is accounted for in the systematic uncertainty of ratios of \Drecoil\ distributions by estimating the systematic uncertainty directly from the spread of the ratios calculated for each variation of the analysis configuration.

%% file: Content/Results.tex
%\clearpage
\section{Results}
\label{Sect:Results}
\subsection{Fully corrected distributions}
%----------------------------
\begin{figure}[htb]
\centering
\includegraphics[width=1.0\textwidth]{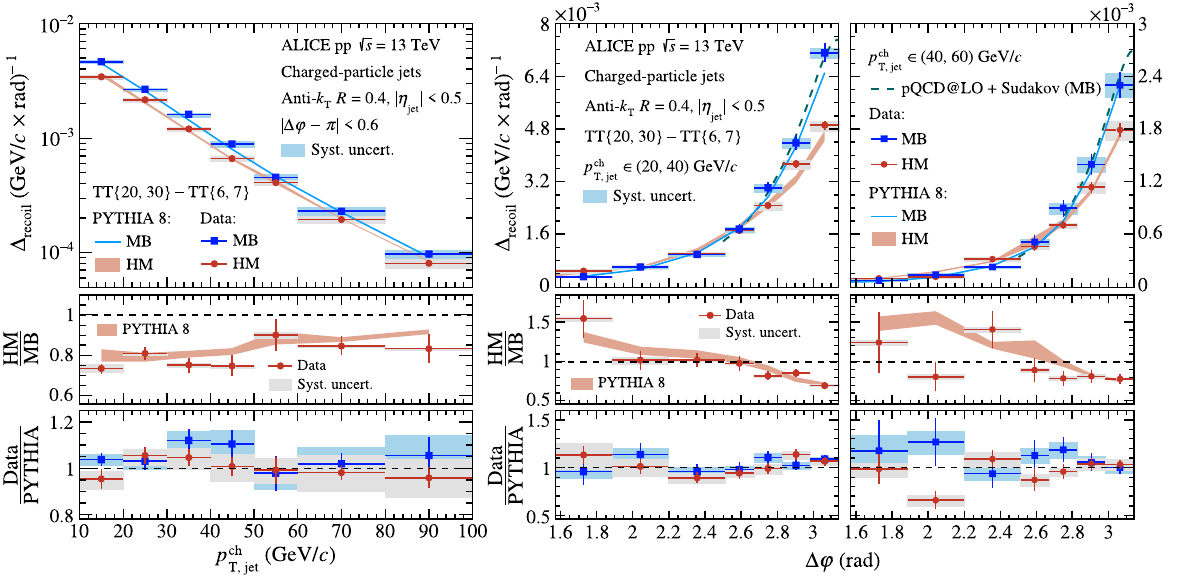}
\caption{Fully-corrected \Drecoil\ distributions measured in MB and HM-selected events in \pp\ collisions at $\sqrts=13$~TeV. Left panel: \Drecoil(\pTjetch) in $|\dphi-\pi|<0.6$; middle and right panels: \Drecoil(\dphi) for $20<\pTjetch<40$~\gev\ and $40<\pTjetch<60$~\gev. Also shown are particle-level simulated distributions calculated with PYTHIA 8 Monash tune, and a pQCD calculation at LO with Sudakov broadening~\cite{Sun:2014gfa,Sun:2015doa,Chen:2016vem} (MB \Drecoil(\dphi) only). The width of the PYTHIA 8 HM band represents the statistical uncertainty. Top row: \Drecoil\ distributions; middle row: ratio of \Drecoil\ distributions for HM/MB from data; bottom row: Data/PYTHIA 8 separately for MB and HM event selections.}
\label{fig:DrecoilSpectraAco}
\end{figure}
%----------------------------

Figure~\ref{fig:DrecoilSpectraAco} shows fully-corrected distributions of \Drecoil(\pTjetch) and \Drecoil(\dphi) measured in MB and HM-selected \pp\ collisions at $\sqrt{s} = 13$ TeV, together with calculations based on PYTHIA~8, Monash tune, and a pQCD calculation at LO with Sudakov broadening~\cite{Sun:2014gfa,Sun:2015doa,Chen:2016vem} (the latter only for \Drecoil(\dphi) in MB collisions). Since the pQCD calculation is LO, there are large uncertainties in its normalization; the \Drecoil(\dphi) distributions from the calculation are therefore scaled to the integrated yield of the data in the same \pTjetch\ bin in order to compare their shapes. Both PYTHIA 8 and the pQCD calculation (\Drecoil(\dphi) shape only) are consistent with the distributions measured in MB events, within experimental uncertainties.

Comparison of the MB and HM \Drecoil(\pTjetch) distributions reveals a yield suppression in HM collisions that is largely independent of \pTjet, though there is a hint of a harder recoil jet spectrum for HM events. The mean value of the yield ratio HM/MB in the left panel of Fig.~\ref{fig:DrecoilSpectraAco} is 0.78. The \Drecoil(\dphi) distributions show that the jet-yield suppression in HM events occurs predominantly in the back-to-back configuration, with the yield ratio HM/MB in the bin $\dphi\sim\pi$ measured to be 0.69 for $20<\pTjetch<40$ \gev\ (Fig.~\ref{fig:DrecoilSpectraAco}, middle panel) and 0.78 for $40<\pTjetch<60$ \gev\ (Fig.~\ref{fig:DrecoilSpectraAco}, left panel). The total yield is suppressed, while the azimuthal distribution is broadened; such broadening could arise from jet quenching, i.e. medium-induced jet scattering occurs preferentially in HM events. Notably, however, PYTHIA 8 particle-level distributions likewise exhibit jet yield suppression and azimuthal broadening for HM-selected events, accurately reproducing the measured distributions. Since PYTHIA 8 does not incorporate jet quenching, this disfavors jet quenching as the predominant effect generating the broadening seen in data.
  
%-----------------------------------------------
\subsection{Origin of HM induced TT--jet acoplanarity  in PYTHIA 8}
%-----------------------------------------------
\begin{figure}[htb]
\centering
\includegraphics[width=1.0\textwidth]{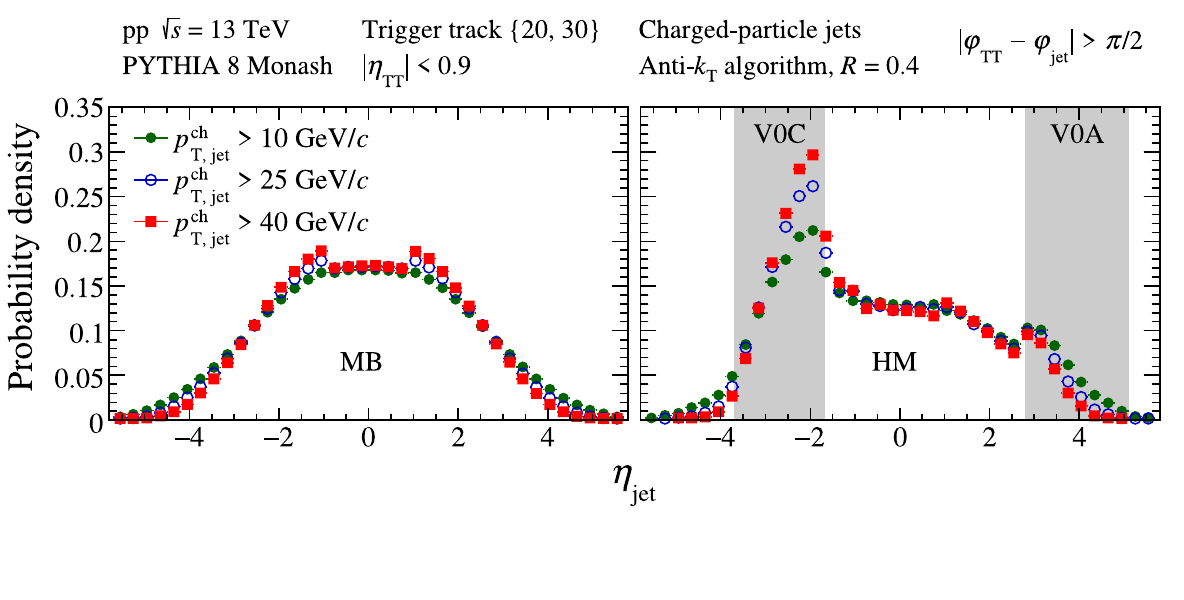}
\caption{PYTHIA 8 particle-level simulation of the probability distribution of the yield of charged-particle jets ($\rr= 0.4$) recoiling from a high-\pT\ hadron (\TT{20}{30}) as a function of \etajet\ for various \pTjetch\ intervals, in \pp\ collisions at $\sqrt{s}=13$~TeV. Left: MB events; right: HM events. V0A and V0C acceptances are also shown.}
\label{Fig:HMbiasA}
\end{figure}
%-----------------------------------------------

To clarify the origin of the broadening, a more detailed investigation is carried out using both data and particle-level distributions from PYTHIA 8 simulations. 

Figure~\ref{Fig:HMbiasA} shows the calculated pseudorapidity (\etajet) distribution of charged-particle jets recoiling from a high-\pT\ trigger hadron with \TT{20}{30} in MB and HM-selected event populations, for various lower thresholds in \pTjetch. Since the per-trigger yield varies with \pTjetch, these distributions are normalized to unit integral to enable a direct comparison of their shapes. The acceptances of V0A and V0C are also shown; note that V0C covers smaller values of $|\eta|$ than V0A. While the \etajet\ distribution for MB events is symmetric, the \etajet\ distribution for HM events is highly asymmetric, with significant enhancement in the relative rate of recoil jets in the V0C acceptance. The enhancement is largest for the highest value of \pTjetch.

%-----------------------------------------------
\begin{figure}[htb]
\centering
\includegraphics[width=0.8\textwidth]{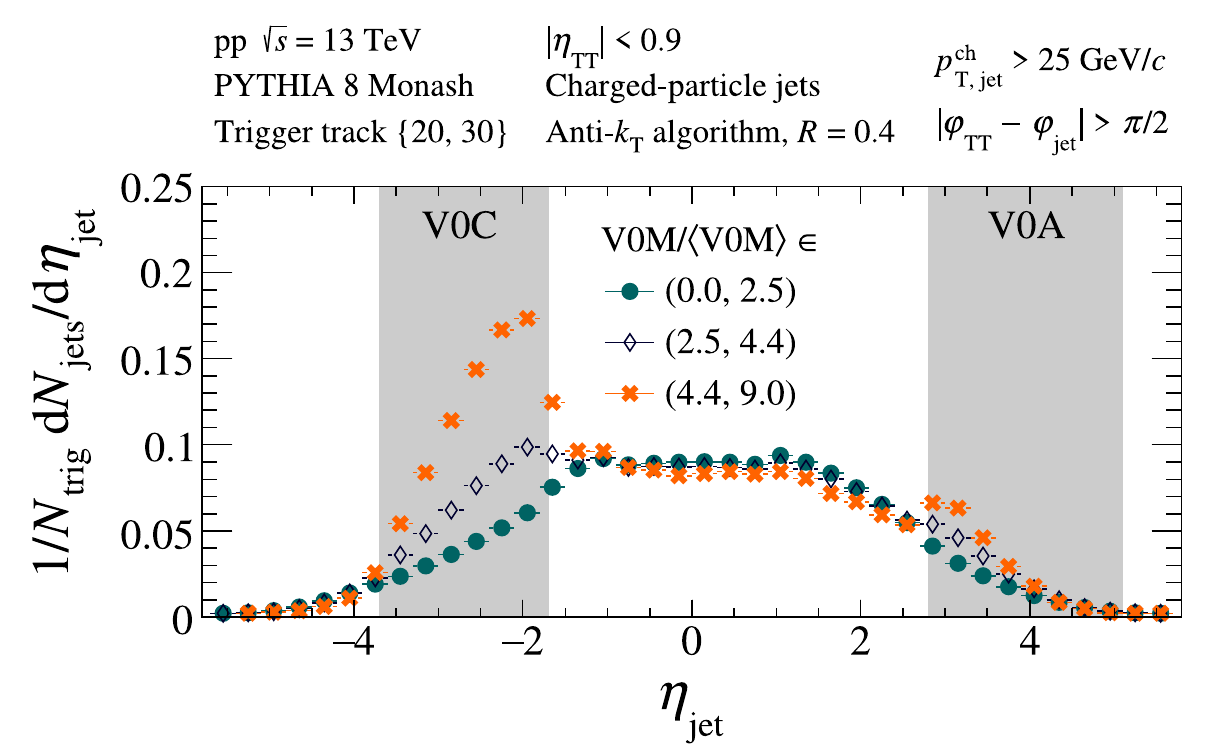}
\caption{Same as Fig.~\ref{Fig:HMbiasA}, but for recoil jets with $\pTjetch>25$~\gev\ for various intervals in \vzeroMscaled. Distributions are normalized per trigger.}
\label{Fig:HMbiasB}
\end{figure}
%-----------------------------------------------

Figure~\ref{Fig:HMbiasB} shows similar distributions for recoil jets with $\pTjetch>25$~\gev, in various intervals of \vzeroMscaled. A striking enhancement is observed in the per-trigger jet yield within the V0C acceptance for the largest values of \vzeroMscaled. Since HM events are selected based on the value of \vzeroMscaled, it is evident that the HM selection induces a bias which enhances the rate of hard recoil jets in the V0C acceptance.

%-----------------------------------------------
\begin{figure}[htb]
\centering
\includegraphics[width=1.0\textwidth]{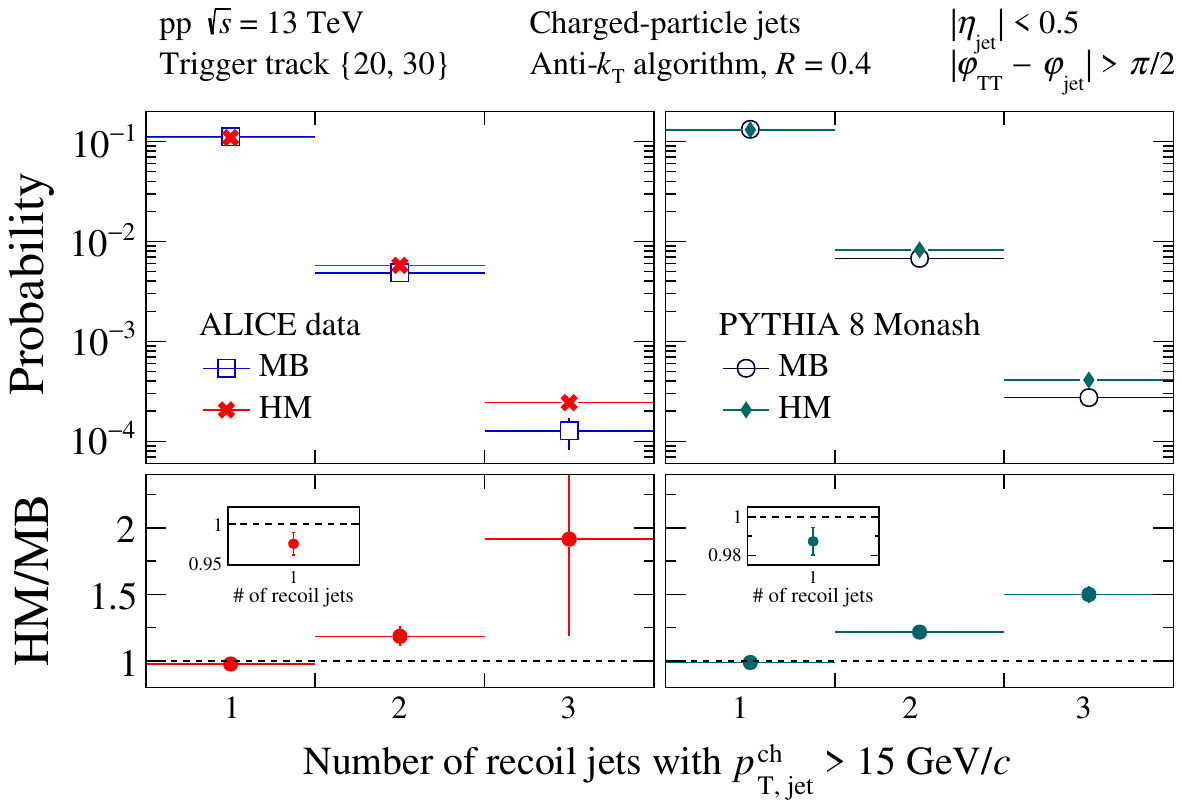}
\caption{Distribution of probability per trigger hadron of number of jets with $\rr=0.4$ and $\pTjetch>15$~\gev\ recoiling from a high-\pT\ hadron (\TT{20}{30}) at midrapidity ($|\etajet|<0.5$), for MB and HM \pp\ collisions at $\sqrts=13$~TeV. Left panels: ALICE data; right panels: simulations at the particle level with PYTHIA 8 Monash tune. Lower panels show the ratio HM/MB. The insert in the lower panels has magnified vertical scale to show the ratio of probabilities to observe a single jet.}
\label{Fig:NjetPythia}
\end{figure}
%-----------------------------------------------

Figure~\ref{Fig:NjetPythia} provides additional insight into the bias induced by the HM event selection. The figure shows the probability to observe a specific number of jets with $\rr=0.4$ and $\pTjetch>15$~\gev\ recoiling against a high-\pT\ hadron trigger (\TT{20}{30}) in the ALICE central barrel acceptance ($|\etajet|<0.5$), for MB and HM \pp\ collisions at $\sqrts=13$ TeV. Distributions are shown for both data and PYTHIA 8 calculations. In both data and simulations, the probability to observe a single jet is suppressed in HM relative to MB events by 2--3\%, while the probability to observe multiple jets is significantly enhanced. 

The effect of the HM selection bias shown in Figs.~\ref{Fig:HMbiasA},~\ref{Fig:HMbiasB} and~\ref{Fig:NjetPythia} on the acoplanarity distributions in Fig.~\ref{fig:DrecoilSpectraAco} can be understood as follows. The HM requirement preferentially selects events with a recoil jet in the V0C acceptance. For the leading-order (LO) di-jet channel, in which the TT and recoil jet are azimuthally back-to-back, this depletes both the per-trigger rate for recoil jets observed in the central barrel ($|\etajet|<0.9-\rr$) at $\dphi\sim\pi$ and the relative rate of single recoil jets in the acceptance. The recoil jets observed in the ALICE central barrel in HM events therefore arise at enhanced rate from higher-order production processes, with multiple recoil jets in the final state. These jets from higher-order processes have broader distribution in \dphi\ than jets from LO production. This picture is confirmed by carrying out the PYTHIA 8 analysis for much larger recoil jet acceptance, $|\etajet|<5.6$. In that case, no back-to-back yield depletion is observed.

The yield suppression and azimuthal broadening seen in Fig.~\ref{fig:DrecoilSpectraAco} may therefore arise from the effect of different phase space being used for EA characterization (V0A and V0C) and recoil jet measurement (central barrel), combined with the HM selection bias towards events with a jet in the V0C acceptance. It therefore cannot be attributed uniquely to jet quenching in HM-selected events.

This conclusion about interpretability in terms of jet quenching of the effects seen in Fig.~\ref{fig:DrecoilSpectraAco} is more general than this specific analysis, however. The bias of the HM event selection identified here is generic, and must also be taken into account for the interpretation in terms of QGP formation of other phenomena observed in small systems.

%-------------------------------------------------
\section{Summary}
\label{Sect:Summary}

This article reports a search for jet quenching effects in high multiplicity \pp\ collisions at $\sqrts=13$ TeV, based on the semi-inclusive azimuthal distribution of charged-particle jets recoiling from a high-\pT\ hadron trigger in the ALICE central barrel acceptance. Significant azimuthal broadening is observed in events selected to have high multiplicity in forward detectors, which may arise from jet quenching. However, similar broadening is also observed in simulations with the PYTHIA 8 event generator, which does not incoporate jet quenching.

Detailed analysis of the data and simulations reveals that the high-multiplicity event selection, which is based on the V0 detectors at forward and backward rapidities, preferentially selects events with an energetic jet in the forward detector, consequently biasing the acoplanarity distribution measured in the central region. We conclude that this observable is therefore not uniquely sensitive to jet quenching effects in small collision systems. This bias is generic, however, and it should likewise be taken into account in the interpretation of other measurements which probe the formation of a quasi-equilibrated quark--gluon plasma in small collision systems.

%% file: fa_2023-08-02_Opt_C.tex
% Version: 2023-08-02

The ALICE Collaboration would like to thank all its engineers and technicians for their invaluable contributions to the construction of the experiment and the CERN accelerator teams for the outstanding performance of the LHC complex.
The ALICE Collaboration gratefully acknowledges the resources and support provided by all Grid centres and the Worldwide LHC Computing Grid (WLCG) collaboration.
The ALICE Collaboration acknowledges the following funding agencies for their support in building and running the ALICE detector:
A. I. Alikhanyan National Science Laboratory (Yerevan Physics Institute) Foundation (ANSL), State Committee of Science and World Federation of Scientists (WFS), Armenia;
Austrian Academy of Sciences, Austrian Science Fund (FWF): [M 2467-N36] and Nationalstiftung f\"{u}r Forschung, Technologie und Entwicklung, Austria;
Ministry of Communications and High Technologies, National Nuclear Research Center, Azerbaijan;
Conselho Nacional de Desenvolvimento Cient\'{\i}fico e Tecnol\'{o}gico (CNPq), Financiadora de Estudos e Projetos (Finep), Funda\c{c}\~{a}o de Amparo \`{a} Pesquisa do Estado de S\~{a}o Paulo (FAPESP) and Universidade Federal do Rio Grande do Sul (UFRGS), Brazil;
Bulgarian Ministry of Education and Science, within the National Roadmap for Research Infrastructures 2020-2027 (object CERN), Bulgaria;
Ministry of Education of China (MOEC) , Ministry of Science \& Technology of China (MSTC) and National Natural Science Foundation of China (NSFC), China;
Ministry of Science and Education and Croatian Science Foundation, Croatia;
Centro de Aplicaciones Tecnol\'{o}gicas y Desarrollo Nuclear (CEADEN), Cubaenerg\'{\i}a, Cuba;
Ministry of Education, Youth and Sports of the Czech Republic, Czech Republic;
The Danish Council for Independent Research | Natural Sciences, the VILLUM FONDEN and Danish National Research Foundation (DNRF), Denmark;
Helsinki Institute of Physics (HIP), Finland;
Commissariat \`{a} l'Energie Atomique (CEA) and Institut National de Physique Nucl\'{e}aire et de Physique des Particules (IN2P3) and Centre National de la Recherche Scientifique (CNRS), France;
Bundesministerium f\"{u}r Bildung und Forschung (BMBF) and GSI Helmholtzzentrum f\"{u}r Schwerionenforschung GmbH, Germany;
General Secretariat for Research and Technology, Ministry of Education, Research and Religions, Greece;
National Research, Development and Innovation Office, Hungary;
Department of Atomic Energy Government of India (DAE), Department of Science and Technology, Government of India (DST), University Grants Commission, Government of India (UGC) and Council of Scientific and Industrial Research (CSIR), India;
National Research and Innovation Agency - BRIN, Indonesia;
Istituto Nazionale di Fisica Nucleare (INFN), Italy;
Japanese Ministry of Education, Culture, Sports, Science and Technology (MEXT) and Japan Society for the Promotion of Science (JSPS) KAKENHI, Japan;
Consejo Nacional de Ciencia (CONACYT) y Tecnolog\'{i}a, through Fondo de Cooperaci\'{o}n Internacional en Ciencia y Tecnolog\'{i}a (FONCICYT) and Direcci\'{o}n General de Asuntos del Personal Academico (DGAPA), Mexico;
Nederlandse Organisatie voor Wetenschappelijk Onderzoek (NWO), Netherlands;
The Research Council of Norway, Norway;
Commission on Science and Technology for Sustainable Development in the South (COMSATS), Pakistan;
Pontificia Universidad Cat\'{o}lica del Per\'{u}, Peru;
Ministry of Education and Science, National Science Centre and WUT ID-UB, Poland;
Korea Institute of Science and Technology Information and National Research Foundation of Korea (NRF), Republic of Korea;
Ministry of Education and Scientific Research, Institute of Atomic Physics, Ministry of Research and Innovation and Institute of Atomic Physics and Universitatea Nationala de Stiinta si Tehnologie Politehnica Bucuresti, Romania;
Ministry of Education, Science, Research and Sport of the Slovak Republic, Slovakia;
National Research Foundation of South Africa, South Africa;
Swedish Research Council (VR) and Knut \& Alice Wallenberg Foundation (KAW), Sweden;
European Organization for Nuclear Research, Switzerland;
Suranaree University of Technology (SUT), National Science and Technology Development Agency (NSTDA) and National Science, Research and Innovation Fund (NSRF via PMU-B B05F650021), Thailand;
Turkish Energy, Nuclear and Mineral Research Agency (TENMAK), Turkey;
National Academy of  Sciences of Ukraine, Ukraine;
Science and Technology Facilities Council (STFC), United Kingdom;
National Science Foundation of the United States of America (NSF) and United States Department of Energy, Office of Nuclear Physics (DOE NP), United States of America.
In addition, individual groups or members have received support from:
Czech Science Foundation (grant no. 23-07499S), Czech Republic;
European Research Council, Strong 2020 - Horizon 2020 (grant nos. 950692, 824093), European Union;
ICSC - Centro Nazionale di Ricerca in High Performance Computing, Big Data and Quantum Computing, European Union - NextGenerationEU;
Academy of Finland (Center of Excellence in Quark Matter) (grant nos. 346327, 346328), Finland.

%% file: 2023-08-02-Alice_Authorlist_2023-08-02_Opt_C.tex
% ALICE Collaboration author list for 2023-08-02
\begin{flushleft} 
\small

S.~Acharya\,\orcidlink{0000-0002-9213-5329}\,$^{\rm 128}$, 
D.~Adamov\'{a}\,\orcidlink{0000-0002-0504-7428}\,$^{\rm 87}$, 
G.~Aglieri Rinella\,\orcidlink{0000-0002-9611-3696}\,$^{\rm 33}$, 
M.~Agnello\,\orcidlink{0000-0002-0760-5075}\,$^{\rm 30}$, 
N.~Agrawal\,\orcidlink{0000-0003-0348-9836}\,$^{\rm 52}$, 
Z.~Ahammed\,\orcidlink{0000-0001-5241-7412}\,$^{\rm 136}$, 
S.~Ahmad\,\orcidlink{0000-0003-0497-5705}\,$^{\rm 16}$, 
S.U.~Ahn\,\orcidlink{0000-0001-8847-489X}\,$^{\rm 72}$, 
I.~Ahuja\,\orcidlink{0000-0002-4417-1392}\,$^{\rm 38}$, 
A.~Akindinov\,\orcidlink{0000-0002-7388-3022}\,$^{\rm 142}$
% ...

\section*{Affiliation Notes}

$^{\rm I}$ Deceased\\
$^{\rm II}$ Also at: Max-Planck-Institut fur Physik, Munich, Germany\\
$^{\rm III}$ Also at: Italian National Agency for New Technologies, Energy and Sustainable Economic Development (ENEA), Bologna, Italy\\
$^{\rm IV}$ Also at: Dipartimento DET del Politecnico di Torino, Turin, Italy\\
$^{\rm V}$ Also at: Yildiz Technical University, Istanbul, T\"{u}rkiye\\
$^{\rm VI}$ Also at: Department of Applied Physics, Aligarh Muslim University, Aligarh, India\\
$^{\rm VII}$ Also at: Institute of Theoretical Physics, University of Wroclaw, Poland\\
$^{\rm VIII}$ Also at: An institution covered by a cooperation agreement with CERN\\

\section*{Collaboration Institutes}

$^{1}$ A.I. Alikhanyan National Science Laboratory (Yerevan Physics Institute) Foundation, Yerevan, Armenia\\
$^{2}$ AGH University of Krakow, Cracow, Poland\\
$^{3}$ Bogolyubov Institute for Theoretical Physics, National Academy of Sciences of Ukraine, Kiev, Ukraine\\
$^{4}$ Bose Institute, Department of Physics  and Centre for Astroparticle Physics and Space Science (CAPSS), Kolkata, India\\
$^{5}$ California Polytechnic State University, San Luis Obispo, California, United States\\
$^{6}$ Central China Normal University, Wuhan, China\\
$^{7}$ Centro de Aplicaciones Tecnol\'{o}gicas y Desarrollo Nuclear (CEADEN), Havana, Cuba\\
$^{8}$ Centro de Investigaci\'{o}n y de Estudios Avanzados (CINVESTAV), Mexico City and M\'{e}rida, Mexico\\
$^{9}$ Chicago State University, Chicago, Illinois, United States\\
$^{10}$ China Institute of Atomic Energy, Beijing, China\\
$^{11}$ China University of Geosciences, Wuhan, China\\
$^{12}$ Chungbuk National University, Cheongju, Republic of Korea\\
$^{13}$ Comenius University Bratislava, Faculty of Mathematics, Physics and Informatics, Bratislava, Slovak Republic\\
$^{14}$ COMSATS University Islamabad, Islamabad, Pakistan\\
$^{15}$ Creighton University, Omaha, Nebraska, United States\\
$^{16}$ Department of Physics, Aligarh Muslim University, Aligarh, India\\
$^{17}$ Department of Physics, Pusan National University, Pusan, Republic of Korea\\
$^{18}$ Department of Physics, Sejong University, Seoul, Republic of Korea\\
$^{19}$ Department of Physics, University of California, Berkeley, California, United States\\
$^{20}$ Department of Physics, University of Oslo, Oslo, Norway\\
$^{21}$ Department of Physics and Technology, University of Bergen, Bergen, Norway\\
$^{22}$ Dipartimento di Fisica, Universit\`{a} di Pavia, Pavia, Italy\\
$^{23}$ Dipartimento di Fisica dell'Universit\`{a} and Sezione INFN, Cagliari, Italy\\
$^{24}$ Dipartimento di Fisica dell'Universit\`{a} and Sezione INFN, Trieste, Italy\\
$^{25}$ Dipartimento di Fisica dell'Universit\`{a} and Sezione INFN, Turin, Italy\\
$^{26}$ Dipartimento di Fisica e Astronomia dell'Universit\`{a} and Sezione INFN, Bologna, Italy\\
$^{27}$ Dipartimento di Fisica e Astronomia dell'Universit\`{a} and Sezione INFN, Catania, Italy\\
$^{28}$ Dipartimento di Fisica e Astronomia dell'Universit\`{a} and Sezione INFN, Padova, Italy\\
$^{29}$ Dipartimento di Fisica `E.R.~Caianiello' dell'Universit\`{a} and Gruppo Collegato INFN, Salerno, Italy\\
$^{30}$ Dipartimento DISAT del Politecnico and Sezione INFN, Turin, Italy\\
$^{31}$ Dipartimento di Scienze MIFT, Universit\`{a} di Messina, Messina, Italy\\
$^{32}$ Dipartimento Interateneo di Fisica `M.~Merlin' and Sezione INFN, Bari, Italy\\
$^{33}$ European Organization for Nuclear Research (CERN), Geneva, Switzerland\\
$^{34}$ Faculty of Electrical Engineering, Mechanical Engineering and Naval Architecture, University of Split, Split, Croatia\\
$^{35}$ Faculty of Engineering and Science, Western Norway University of Applied Sciences, Bergen, Norway\\
$^{36}$ Faculty of Nuclear Sciences and Physical Engineering, Czech Technical University in Prague, Prague, Czech Republic\\
$^{37}$ Faculty of Physics, Sofia University, Sofia, Bulgaria\\
$^{38}$ Faculty of Science, P.J.~\v{S}af\'{a}rik University, Ko\v{s}ice, Slovak Republic\\
$^{39}$ Frankfurt Institute for Advanced Studies, Johann Wolfgang Goethe-Universit\"{a}t Frankfurt, Frankfurt, Germany\\
$^{40}$ Fudan University, Shanghai, China\\
$^{41}$ Gangneung-Wonju National University, Gangneung, Republic of Korea\\
$^{42}$ Gauhati University, Department of Physics, Guwahati, India\\
$^{43}$ Helmholtz-Institut f\"{u}r Strahlen- und Kernphysik, Rheinische Friedrich-Wilhelms-Universit\"{a}t Bonn, Bonn, Germany\\
$^{44}$ Helsinki Institute of Physics (HIP), Helsinki, Finland\\
$^{45}$ High Energy Physics Group,  Universidad Aut\'{o}noma de Puebla, Puebla, Mexico\\
$^{46}$ Horia Hulubei National Institute of Physics and Nuclear Engineering, Bucharest, Romania\\
$^{47}$ HUN-REN Wigner Research Centre for Physics, Budapest, Hungary\\
$^{48}$ Indian Institute of Technology Bombay (IIT), Mumbai, India\\
$^{49}$ Indian Institute of Technology Indore, Indore, India\\
$^{50}$ INFN, Laboratori Nazionali di Frascati, Frascati, Italy\\
$^{51}$ INFN, Sezione di Bari, Bari, Italy\\
$^{52}$ INFN, Sezione di Bologna, Bologna, Italy\\
$^{53}$ INFN, Sezione di Cagliari, Cagliari, Italy\\
$^{54}$ INFN, Sezione di Catania, Catania, Italy\\
$^{55}$ INFN, Sezione di Padova, Padova, Italy\\
$^{56}$ INFN, Sezione di Pavia, Pavia, Italy\\
$^{57}$ INFN, Sezione di Torino, Turin, Italy\\
$^{58}$ INFN, Sezione di Trieste, Trieste, Italy\\
$^{59}$ Inha University, Incheon, Republic of Korea\\
$^{60}$ Institute for Gravitational and Subatomic Physics (GRASP), Utrecht University/Nikhef, Utrecht, Netherlands\\
$^{61}$ Institute of Experimental Physics, Slovak Academy of Sciences, Ko\v{s}ice, Slovak Republic\\
$^{62}$ Institute of Physics, Homi Bhabha National Institute, Bhubaneswar, India\\
$^{63}$ Institute of Physics of the Czech Academy of Sciences, Prague, Czech Republic\\
$^{64}$ Institute of Space Science (ISS), Bucharest, Romania\\
$^{65}$ Institut f\"{u}r Kernphysik, Johann Wolfgang Goethe-Universit\"{a}t Frankfurt, Frankfurt, Germany\\
$^{66}$ Instituto de Ciencias Nucleares, Universidad Nacional Aut\'{o}noma de M\'{e}xico, Mexico City, Mexico\\
$^{67}$ Instituto de F\'{i}sica, Universidade Federal do Rio Grande do Sul (UFRGS), Porto Alegre, Brazil\\
$^{68}$ Instituto de F\'{\i}sica, Universidad Nacional Aut\'{o}noma de M\'{e}xico, Mexico City, Mexico\\
$^{69}$ iThemba LABS, National Research Foundation, Somerset West, South Africa\\
$^{70}$ Jeonbuk National University, Jeonju, Republic of Korea\\
$^{71}$ Johann-Wolfgang-Goethe Universit\"{a}t Frankfurt Institut f\"{u}r Informatik, Fachbereich Informatik und Mathematik, Frankfurt, Germany\\
$^{72}$ Korea Institute of Science and Technology Information, Daejeon, Republic of Korea\\
$^{73}$ KTO Karatay University, Konya, Turkey\\
$^{74}$ Laboratoire de Physique Subatomique et de Cosmologie, Universit\'{e} Grenoble-Alpes, CNRS-IN2P3, Grenoble, France\\
$^{75}$ Lawrence Berkeley National Laboratory, Berkeley, California, United States\\
$^{76}$ Lund University Department of Physics, Division of Particle Physics, Lund, Sweden\\
$^{77}$ Nagasaki Institute of Applied Science, Nagasaki, Japan\\
$^{78}$ Nara Women{'}s University (NWU), Nara, Japan\\
$^{79}$ National and Kapodistrian University of Athens, School of Science, Department of Physics , Athens, Greece\\
$^{80}$ National Centre for Nuclear Research, Warsaw, Poland\\
$^{81}$ National Institute of Science Education and Research, Homi Bhabha National Institute, Jatni, India\\
$^{82}$ National Nuclear Research Center, Baku, Azerbaijan\\
$^{83}$ National Research and Innovation Agency - BRIN, Jakarta, Indonesia\\
$^{84}$ Niels Bohr Institute, University of Copenhagen, Copenhagen, Denmark\\
$^{85}$ Nikhef, National institute for subatomic physics, Amsterdam, Netherlands\\
$^{86}$ Nuclear Physics Group, STFC Daresbury Laboratory, Daresbury, United Kingdom\\
$^{87}$ Nuclear Physics Institute of the Czech Academy of Sciences, Husinec-\v{R}e\v{z}, Czech Republic\\
$^{88}$ Oak Ridge National Laboratory, Oak Ridge, Tennessee, United States\\
$^{89}$ Ohio State University, Columbus, Ohio, United States\\
$^{90}$ Physics department, Faculty of science, University of Zagreb, Zagreb, Croatia\\
$^{91}$ Physics Department, Panjab University, Chandigarh, India\\
$^{92}$ Physics Department, University of Jammu, Jammu, India\\
$^{93}$ Physics Program and International Institute for Sustainability with Knotted Chiral Meta Matter (SKCM2), Hiroshima University, Hiroshima, Japan\\
$^{94}$ Physikalisches Institut, Eberhard-Karls-Universit\"{a}t T\"{u}bingen, T\"{u}bingen, Germany\\
$^{95}$ Physikalisches Institut, Ruprecht-Karls-Universit\"{a}t Heidelberg, Heidelberg, Germany\\
$^{96}$ Physik Department, Technische Universit\"{a}t M\"{u}nchen, Munich, Germany\\
$^{97}$ Politecnico di Bari and Sezione INFN, Bari, Italy\\
$^{98}$ Research Division and ExtreMe Matter Institute EMMI, GSI Helmholtzzentrum f\"ur Schwerionenforschung GmbH, Darmstadt, Germany\\
$^{99}$ Saga University, Saga, Japan\\
$^{100}$ Saha Institute of Nuclear Physics, Homi Bhabha National Institute, Kolkata, India\\
$^{101}$ School of Physics and Astronomy, University of Birmingham, Birmingham, United Kingdom\\
$^{102}$ Secci\'{o}n F\'{\i}sica, Departamento de Ciencias, Pontificia Universidad Cat\'{o}lica del Per\'{u}, Lima, Peru\\
$^{103}$ Stefan Meyer Institut f\"{u}r Subatomare Physik (SMI), Vienna, Austria\\
$^{104}$ SUBATECH, IMT Atlantique, Nantes Universit\'{e}, CNRS-IN2P3, Nantes, France\\
$^{105}$ Sungkyunkwan University, Suwon City, Republic of Korea\\
$^{106}$ Suranaree University of Technology, Nakhon Ratchasima, Thailand\\
$^{107}$ Technical University of Ko\v{s}ice, Ko\v{s}ice, Slovak Republic\\
$^{108}$ The Henryk Niewodniczanski Institute of Nuclear Physics, Polish Academy of Sciences, Cracow, Poland\\
$^{109}$ The University of Texas at Austin, Austin, Texas, United States\\
$^{110}$ Universidad Aut\'{o}noma de Sinaloa, Culiac\'{a}n, Mexico\\
$^{111}$ Universidade de S\~{a}o Paulo (USP), S\~{a}o Paulo, Brazil\\
$^{112}$ Universidade Estadual de Campinas (UNICAMP), Campinas, Brazil\\
$^{113}$ Universidade Federal do ABC, Santo Andre, Brazil\\
$^{114}$ Universitatea Nationala de Stiinta si Tehnologie Politehnica Bucuresti, Bucharest, Romania\\
$^{115}$ University of Cape Town, Cape Town, South Africa\\
$^{116}$ University of Derby, Derby, United Kingdom\\
$^{117}$ University of Houston, Houston, Texas, United States\\
$^{118}$ University of Jyv\"{a}skyl\"{a}, Jyv\"{a}skyl\"{a}, Finland\\
$^{119}$ University of Kansas, Lawrence, Kansas, United States\\
$^{120}$ University of Liverpool, Liverpool, United Kingdom\\
$^{121}$ University of Science and Technology of China, Hefei, China\\
$^{122}$ University of South-Eastern Norway, Kongsberg, Norway\\
$^{123}$ University of Tennessee, Knoxville, Tennessee, United States\\
$^{124}$ University of the Witwatersrand, Johannesburg, South Africa\\
$^{125}$ University of Tokyo, Tokyo, Japan\\
$^{126}$ University of Tsukuba, Tsukuba, Japan\\
$^{127}$ Universit\"{a}t M\"{u}nster, Institut f\"{u}r Kernphysik, M\"{u}nster, Germany\\
$^{128}$ Universit\'{e} Clermont Auvergne, CNRS/IN2P3, LPC, Clermont-Ferrand, France\\
$^{129}$ Universit\'{e} de Lyon, CNRS/IN2P3, Institut de Physique des 2 Infinis de Lyon, Lyon, France\\
$^{130}$ Universit\'{e} de Strasbourg, CNRS, IPHC UMR 7178, F-67000 Strasbourg, France, Strasbourg, France\\
$^{131}$ Universit\'{e} Paris-Saclay, Centre d'Etudes de Saclay (CEA), IRFU, D\'{e}partment de Physique Nucl\'{e}aire (DPhN), Saclay, France\\
$^{132}$ Universit\'{e}  Paris-Saclay, CNRS/IN2P3, IJCLab, Orsay, France\\
$^{133}$ Universit\`{a} degli Studi di Foggia, Foggia, Italy\\
$^{134}$ Universit\`{a} del Piemonte Orientale, Vercelli, Italy\\
$^{135}$ Universit\`{a} di Brescia, Brescia, Italy\\
$^{136}$ Variable Energy Cyclotron Centre, Homi Bhabha National Institute, Kolkata, India\\
$^{137}$ Warsaw University of Technology, Warsaw, Poland\\
$^{138}$ Wayne State University, Detroit, Michigan, United States\\
$^{139}$ Yale University, New Haven, Connecticut, United States\\
$^{140}$ Yonsei University, Seoul, Republic of Korea\\
$^{141}$  Zentrum  f\"{u}r Technologie und Transfer (ZTT), Worms, Germany\\
$^{142}$ Affiliated with an institute covered by a cooperation agreement with CERN\\
$^{143}$ Affiliated with an international laboratory covered by a cooperation agreement with CERN.\\

\end{flushleft} 